\documentclass[preprint,5p,12pt]{elsarticle}
\usepackage{lineno,hyperref}
\modulolinenumbers[5]

\journal{Journal of \LaTeX\ Templates}

\usepackage{epsfig}
\usepackage{supertabular}
\usepackage{longtable}
\usepackage{graphicx}
\usepackage[figuresright]{rotating}
\journal{Physics Letters B}

\begin{document}

\begin{frontmatter}

  \title{The microscopic mechanism behind the fission-barrier asymmetry (II):\\
         The rare-earth region $50 < Z < 82$ and $82 < N < 126$. }

\author{T. Ichikawa}
\address{Yukawa Institute for Theoretical Physics, Kyoto University, Kyoto
606-8502, Japan}

\author{P. M{\"{o}}ller}
\address{Theoretical Division, Los Alamos National Laboratory, Los Alamos, NM
  87544, USA}
\address{P. Moller Scientific Computing and Graphics, Inc., P. O. Box 1440, Los Alamos, NM 87544, USA}
\cortext[mycorrespondingauthor]{Corresponding author}
\ead{mollerinla@gmail.com}
















\begin{abstract}
  It is well known that most actinides fission into fragments of unequal
  size. This contradicts  liquid-drop-model
  theory  from which symmetric fission is expected.
  The first attempt to understand this difference
  suggested that division leading to one of the 
  fragments being near doubly magic $^{132}$Sn is favored by gain in
  binding energy.
  After the Strutinsky shell-correction method was developed
  an alternative idea that gained popularity was that the fission saddle might
  be lower for mass-asymmetric shapes and that this asymmetry
  was preserved until scission. Recently it
  was observed [Phys. Rev. Lett. {\bf 105} (2010) 252502]
  that $^{180}$Hg preferentially fissions asymmetrically in
  contradiction to the fragment-magic-shell expectation which suggested
  symmetric division peaked around $^{90}$Zr, with its magic neutron
  number $N=50$, so it was presented as
  a ``new type of asymmetric fission''.
  However, in a paper [Phys. Lett. 34B (1971) 349]
  a ``simple'' microscopic mechanism 
  behind the asymmetry of the actinide fission saddle points
  was  proposed to be related
  the  coupling between levels of type [40$\Lambda\Omega$]
  and [51$\Lambda\Omega$]. 
  The paper then generalizes this idea and made the remarkable prediction 
  that analogous features could exist in other regions. In particular
  it was proposed that
  in the rare-earth region couplings between levels of type [30$\Lambda\Omega$]
  and [41$\Lambda\Omega$] would favor mass-asymmetric outer saddle shapes.
  In this picture the asymmetry of $^{180}$Hg is not a ``new type of
  asymmetric fission'' but of analogous  origin as the asymmetry of
  actinide fission. This prediction has never been cited
  in the discussion of the recently observed fission asymmetries in
  the ``new region of asymmetry'', in nuclear physics also referred
  to as the rare-earth region.
  We show by detailed analysis that the mechanism of the saddle asymmetry
  in the sub-Pb region is indeed the one predicted half a century ago.
\end{abstract}

\begin{keyword}
\texttt{elsarticle.cls}\sep \LaTeX\sep Elsevier \sep template
\MSC[2010] 00-01\sep  99-00
\end{keyword}

\end{frontmatter}
\section{Introduction}
The discovery of fission in 1938 was based on the identification of
barium ($Z=55)$ in the products following bombardment of uranium with
neutrons \cite{hahn39:a}.  An immediate intuitive theoretical model
providing a picture of the phenomenon in terms of the deformation of a
charged liquid drop with a surface tension was given by Meitner and
Frisch \cite{meitner39:a}. The discovery and its interpretation was
further confirmed by observation of the high kinetic energies of the
fission fragments \cite{frisch39:a}.  About half a year later Bohr and
Wheeler provided a more complete theoretical and quantitative  discussion of the
observed fission process by generalizing the semiempirical mass model
\cite{bethe36:a} into a liquid-drop model of the nuclear potential
energy as a function shape \cite{bohr39:a}. 

However, the liquid-drop model theory did not explain the observations
that the preferred mass split, of the light actinide systems
studied at the time, was asymmetric mass division with
a heavy fragment with nucleon number $A \approx 140$ and the remaining
nucleons in a smaller fragment. The energetically preferred division in
liquid-drop model theory is symmetric. Since the discovery of fission
a subject of intense interest has been and still is to explain the observed
fission asymmetry and ideally to model more exactly
the observed yield distributions.

An initial qualitative theoretical interpretation for the experimental
observations of asymmetric fission was that fissioning systems favor
division into a heavy fragment near the doubly magic $^{132}$Sn
because the magic proton number $Z=50$ and neutron number
$N=82$ and associated microscopic effects result in an extra binding of
about 12 MeV in$^{132}$Sn relative to liquid-drop
theory.  In this interpretation asymmetric fission would be
roughly limited to the actinide region because no nuclei outside
this region can divide into a doubly magic fragment while preserving
the $Z/N$ of the fissioning nucleus, a necessity due to the rapid
increase in the symmetry energy for deviations from this ratio.

Once, after Strutinsky had introduced his quantitative method
\cite{strutinsky67:a,strutinsky68:a} for calculating how microscopic
effects lead to differences in the calculated nuclear potential
energies, relative to the liquid drop model, another explanation for
asymmetric fission was investigated: asymmetric fission occurs because
the lowest saddle between the ground-state and separated fragments
would correspond to mass-asymmetric shapes.  The first calculations
showing that actinide shapes at the energetically most favorable
saddle correspond to mass-asymmetric shapes and for lighter systems
correspond to symmetric shapes were obtained in 1970
\cite{moller70:a}.  These results were confirmed a year later by
another group \cite{pauli71:a}, also in a Strutinsky-type calculation
but with a different single-particle model (Woods-Saxon), so the
results appeared very robust. Only limited studies of the occurrence
of asymmetric saddle point shapes outside the actinide region were
performed at this time.  It was implicitly assumed that the shape
asymmetry at the saddle was preserved until the final fragments formed
and separated. We will discuss why and to what degree this is the case
in many fissioning nuclei. Early  papers did indeed show a strong
correlation between the calculated degree of asymmetry of the nuclear
shape at the saddle point and the observed final mass asymmetry, see for
example Refs.\ \cite{pauli71:a,moller72:a}.
These early studies of the asymmetry of the saddle points were based
on studies of the nuclear potential energy  as a function of a very few
shapes, 175 in a 1974 calculation \cite{moller74:a,moller74:b} for example.
It was generally recognized that
more realistic models of fission fragment asymmetries needed to
add dynamical or statistical considerations in addition to
potential-energy calculations, and obviously also
include microscopic effects. A vast number of various types of
such calculations have been carried out over they years, but partly because
of limited computer power many approximations were made. In particular
the potential energy was not calculated for a sufficiently large
number of nuclear shapes and no model that became generally used
emerged until very recently.
Also, models were not applied to studies of large regions
of nuclei, so it was impossible to understand
if they were sufficiently realistic to {\it predict}  fission-fragment yields.
In fact, when K.-H. Schmidt performed his seminal study
of fission-fragment charge distributions of 70 different fissioning species
\cite{schmidt00:a}
and tried to find corresponding calculations that showed the transition
region between symmetric and asymmetric fission that he observed
in the regions he studied ($ 85\leq Z \leq 94$) he found only a 30-year old,
very simple study that covered the region of interest \cite{moller72:a}.

\section{Current status of fission-fragment mass-distribution studies}
It has been argued that to calculate a sufficiently general
nuclear potential-energy surface it must be a function of all major shape
types that the nucleus exploits during the transition from a single
parent nucleus to separated fragments, namely elongation, neck size,
the two (independent) nascent fragment shapes, and mass asymmetry,
that is five independently variable
shape parameters (referred to as a ``5D'' calculation because
of the five independent shape variables).
Such a calculation, which should be a
continuous function of the shape variables to be meaningful,
makes it necessary to calculate the potential energy for several
million different nuclear shapes. The necessary computer power
has only been available for the last 20 years or so. The first such
calculations are discussed in Refs.\ \cite{moller99:a,moller00:a,moller01:a}.
More complete discussions and extensions to more than 5000 nuclei
in the regions $171\leq A \leq 330$ are in Refs.\ \cite{moller09:a,moller15:a}.
The most probable mass split was again determined from the structure
of the potential-energy surface, now somewhat refined so the
asymmetry was obtained from the asymmetry of the
``asymmetric fission valley'' as discussed in Ref.\ \cite{moller01:a}.
The results agreed with the most probable observed mass splits
to within 3 nucleons.

Around the year 2000 there still existed  only a
few experimental studies of fission-fragment
mass distributions substantially  below the actinide region. A study by Itkis
\cite{itkis90:a,itkis91:a}
of fission-fragment distributions of nuclides in the region $A\approx 200$
was interpreted as showing a hint of mass asymmetry at excitation energies
of about 30 MeV, or 10 MeV above the saddle point.

A game changer in experimental fission  studies was that of the
$\beta$-delayed fission of  $^{180}$Tl in
Ref.\ \cite{andreyev10:a}. It is the daughter $^{180}$Hg following
the $\beta$-decay of  $^{180}$Tl  that fissions.
The excitation energy cannot exceed
the $Q$-value of the $\beta$ decay, slightly over 10 MeV and just barely
above the fission saddle energy so microscopic effects can be
expected to be much more expressed than in the studies by Itkis.
It is stated in the abstract of Ref.\ \cite{andreyev10:a}
that common expectations at the time of the experiment
were that fission would be symmetric
because it would lead to two $^{90}$Zn fragments with $N=50$ magic
and $Z=40$ semimagic. However, the experiment established  100/80
as the  most probable
mass split. It was also
stated it was a new type of asymmetric fission because the
observed mass split was not related
to fragment shell effects. A 5D potential-energy calculation based on
the model in \cite{moller01:a,moller09:a} and included
in Ref.\ \cite{andreyev10:a} found a saddle asymmetry
of 108/72.
\begin{figure*}[t!]
  \includegraphics[width=\textwidth]{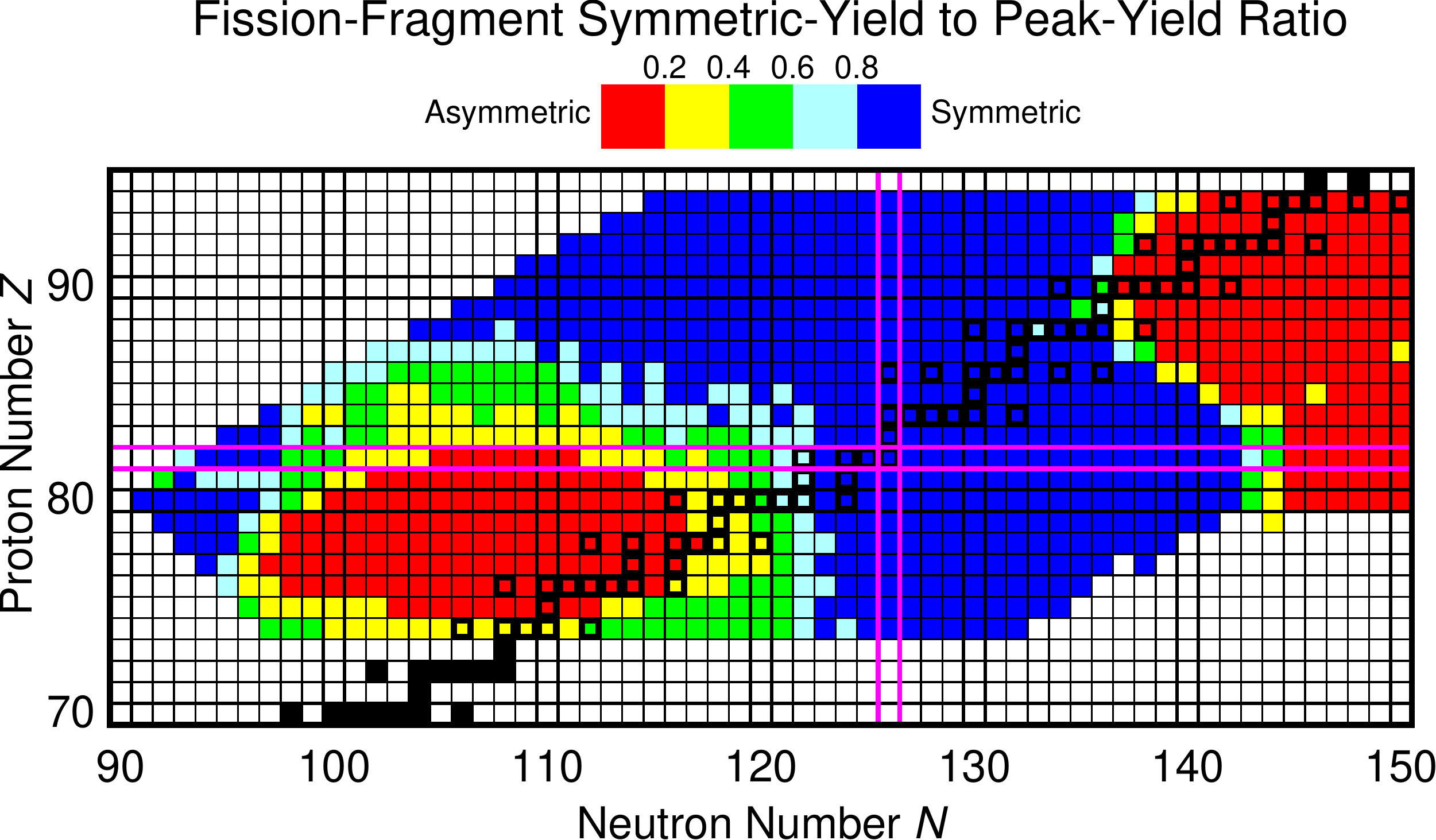}
  \caption{ Calculated symmetric-yield to peak-yield ratios for 987 fissioning systems. Black squares (open in colored regions,
filled outside) indicate $\beta$-stable nuclei. From Ref. \protect\cite{moller15:b} where the results in the figure are further discussed.}
\label{asymreg}
\end{figure*}

A second, now theoretical, game changer took place a few years later.
It was for the first time shown in Ref.\ \cite{randrup11:a}
that fission-fragment
mass distributions could  be routinely, quantitatively
and globally calculated, with an accuracy not previously
achieved, in a consistent and  well-defined model.
In this approach ``dynamical'' aspects
of the evolution from ground state to separated fragments
are treated as  random walks on the previously
calculated 5D potential-energy surfaces, with the walk simulated by means
of the Metropolis procedure.  In its first implementation no parameters were
adjusted to experimental mass yields. Still, excellent agreement with the
eight test cases studied in \cite{randrup11:a} was achieved. Later a
phenomenological damping of the shell correction with two globally
adjusted parameters was included \cite{randrup13:a}, in which paper
yields were calculated and compared to the 70 charge yields measured
by Ref.\ \cite{schmidt00:a}, with excellent agreement.
The model was soon after used to predict fission yields for 987 nuclides
and the most probable fragment mass asymmetry
in the regions $74 \leq Z \leq 94$ \cite{moller15:b}. We reproduce a figure
from that paper as Fig.\ \ref{asymreg} here.

\section{The microscopic mechanism behind the fission saddle asymmetry}

Very soon after the first calculations \cite{moller70:a,pauli71:a}
of fission saddle point asymmetries by use
of the Strutinsky shell-correction method, 
an explanation, 
in simple terms, of {\it why} the second saddles in actinides correspond to
asymmetric  shapes was presented in Ref.\ \cite{gustafsson71:a}.
As is common in nuclear physics in discussions of single-particle
levels and transition rates the concept and notation
of asymptotic quantum numbers [$N\; n_{\rm z}\; \Lambda \; \Omega$] is
used in this reference as we therefore
also  do here. These are the quantum numbers
of the eigenfunctions of a deformed,
axially symmetric harmonic
oscillator \cite{mottelson59:a,copley66:a,damgaard69:a}.
At first sight it might not seem possible to give a simple explanation
for why some saddle shapes are asymmetric because
the potential energy at a specific shape, such as the saddle,
is a sum of macroscopic Coulomb- and surface-energy
terms, each about 800 MeV (where the leading-order term in a deformation 
expansion in terms of $\beta_{2}$ is of different
sign in the Coulomb and surface energy terms) and a microscopic shell-correction
term of magnitude $\pm 10$ MeV. Each of these terms depends in a complicated
way on the several deformation parameters considered, with large cancellations
occurring between the shape-dependent parts of
the Coulomb- and surface-energy terms. In addition
the shell corrections depend on level structure. Therefore, to arrive
at a ``simple'' explanation for the asymmetry {\it at the saddle point}
might seem impossible, because variations in the sum of the
above terms of only a few MeV determine if the saddle shape is
symmetric or asymmetric.

\subsection{The microscopic origin of
  the asymmetry of the fission saddle in actinide nuclei}
Ref.\ \cite{gustafsson71:a} approached the issue of why
the saddle point is asymmetric by avoiding the difficult issue of
the complicated behavior of the multiterm, multivariable
potential-energy function. It approached the question from
a different direction by  showing
that the key reason for asymmetric saddle points in actinide nuclei is
the influence of neutron orbitals of types [40$\Lambda \Omega$] 
and [51$\Lambda \Omega$], in particular the interaction between
these levels.
The latter are for small elongations
above the former but slope less than the former
with increasing elongation so the
two types become close at the elongated
saddle distortions with the neutron Fermi surface
roughly in between the two level types. See Fig.\ 2 in Ref.\
\cite{gustafsson71:a} for the behavior of these levels with
deformation. When the nucleus is asymmetrically deformed at the saddle point
the [51$\Lambda \Omega$] levels bend upwards an the [40$\Lambda \Omega$] levels
bend downwards, due to couplings, or matrix elements,
between these wave functions via the asymmetry part of
the potential as schematically discussed and shown
in Fig.\ 3 in Ref.\ \cite{gustafsson71:a}.
\begin{figure}[t!]
  \includegraphics[width=\linewidth]{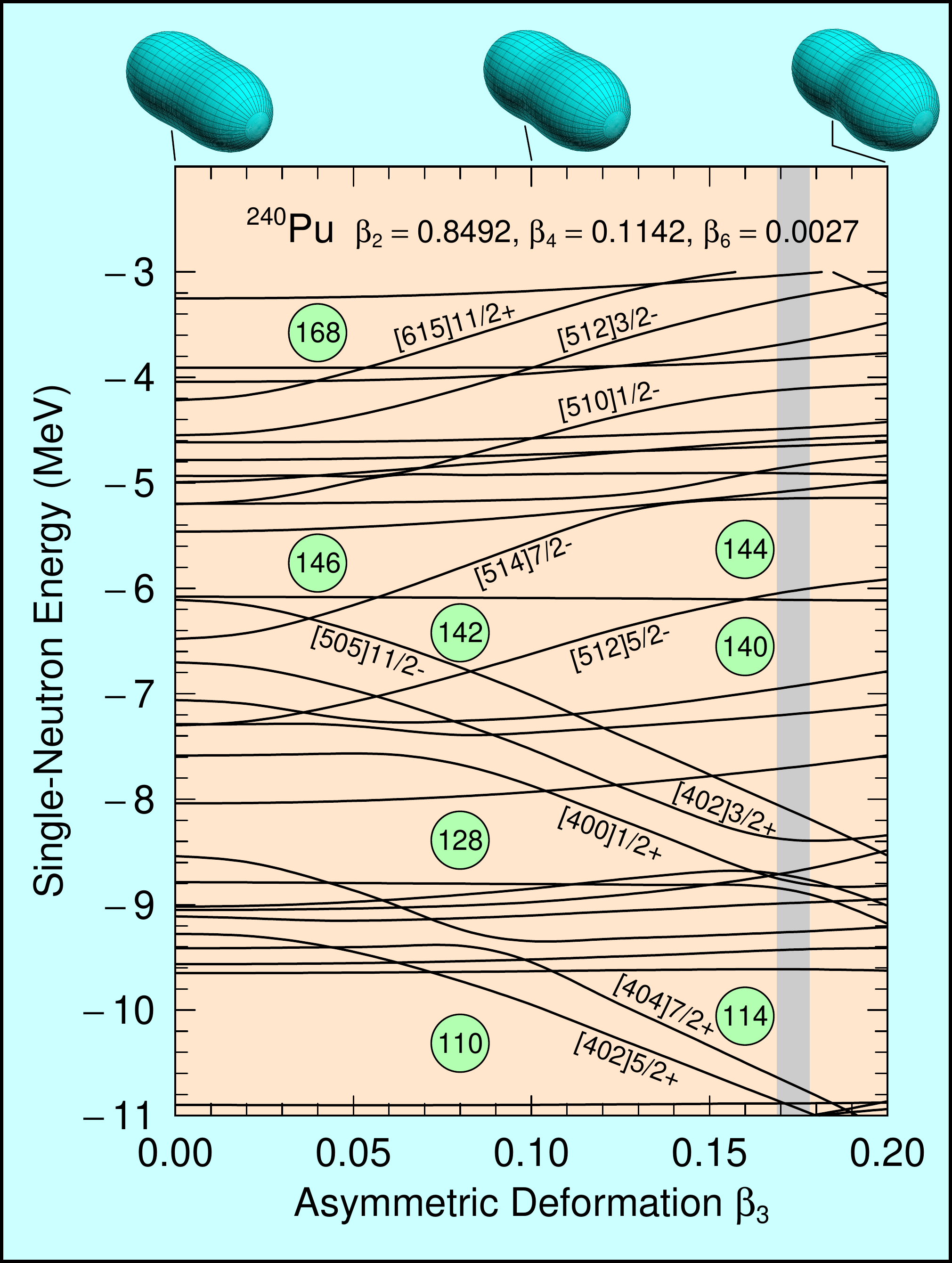}
  \caption{Levels versus the asymmetry coordinate $\beta_3$ starting
    from the outer {\it symmetric} saddle of $^{240}$Pu.
    The level behavior
    closely resembles Fig.\ 3 in Ref.\ \protect\cite{gustafsson71:a},
    illustrating
    the microscopic mechanism behind the actinide saddle asymmetry,
    as explained half a century ago. The gray
    vertical column is located at the calculated
    saddle asymmetry. For clarity we have here placed $\Omega$ to the right of
  the right square bracket, in contrast to in the text.}
\label{240pulev}
\end{figure}

\begin{figure}[t!]
  \includegraphics[width=\linewidth]{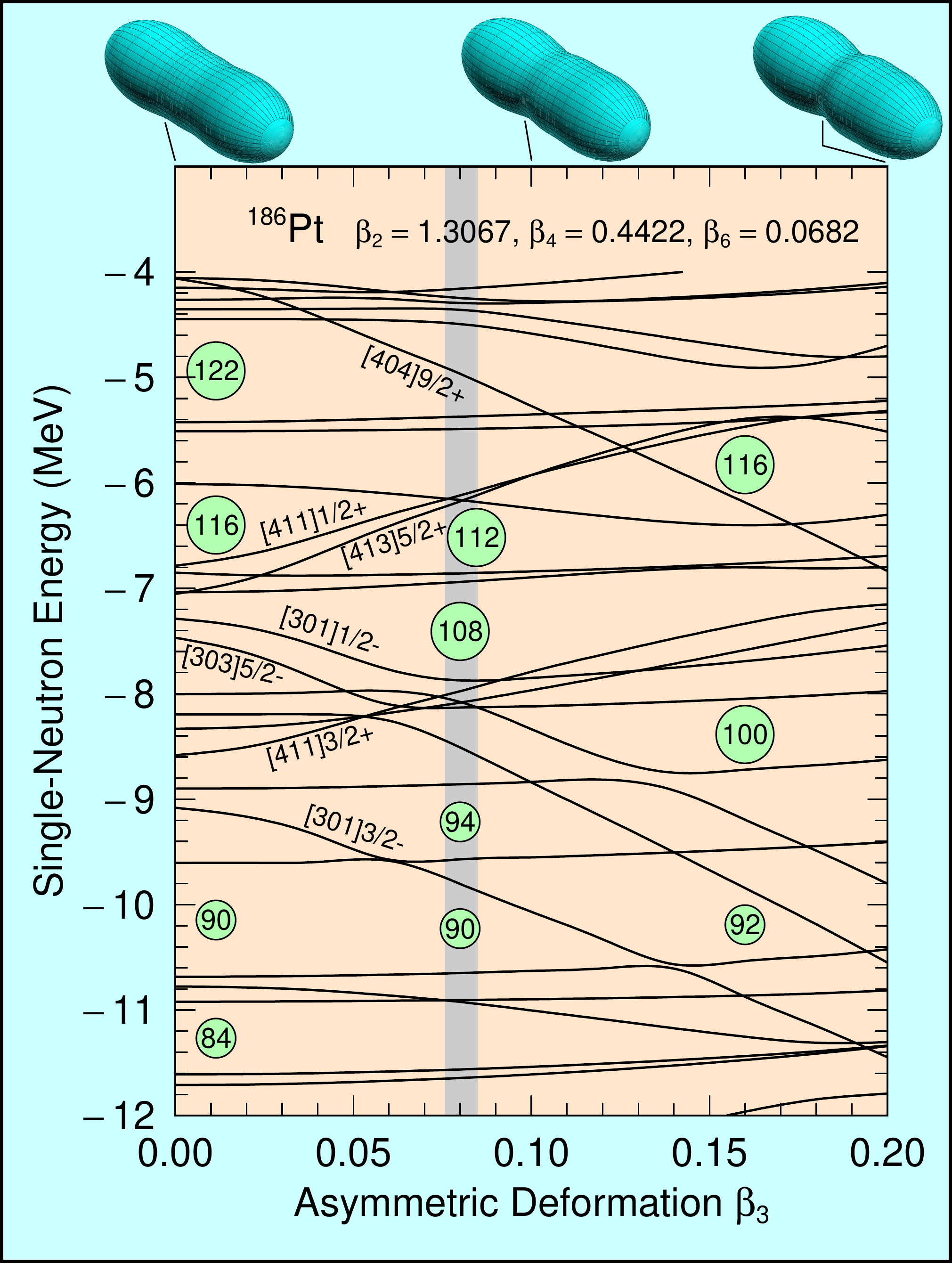}
  \caption[capt01]{Levels versus the asymmetry coordinate $\beta_3$ starting
    from the symmetric saddle of $^{186}_{\phantom{0}78}$Pt. Just as predicted
  half a century ago in Ref.\ \protect\cite{gustafsson71:a}
  the levels  of type 
   [30$\Lambda \Omega$]  couple strongly to levels of type [41$\Lambda \Omega$]
   leading to a large gap for asymmetric shapes at the neutron Fermi surface
   ($N=108$) and therefore
   asymmetric saddle shapes are energetically preferred. The gray
   vertical column is
   located at the calculated saddle asymmetry.
  For clarity we have here placed $\Omega$ to the right of
  the right square bracket, in contrast to in the text.}
\label{186ptlev}
\end{figure}
We show here the corresponding results in our most current model
in Fig.\ \ref{240pulev}. The saddle points have been calculated
in our full 5-dimensional three-quadratic-surface parameterization as
described in Ref.\ \cite{moller09:a}. However, one-dimensional
level diagrams are most
suitably displayed in less ``non-linear'' coordinates
than those of the three-quadratic-surface parameterization. Therefore we have
converted the three-quadratic-surface shape coordinates at the saddle points to
$\beta$ shape coordinates using Eq.\ 2 in Ref.\ \cite{moller09:a}.
The levels start at the calculated symmetric saddle point at
$\beta_{3}$ = 0. As $\beta_{3}$ increases we see a very similar behavior
as in Ref.\ \cite{gustafsson71:a}. This is despite the
vastly smaller deformation space, about 100 different shapes, used in
Ref.\ \cite{gustafsson71:a} to determine saddle-point shapes, compared to the millions of shapes included
in Ref.\ \cite{moller09:a}.
Remarkably the level order within
the level groups [40$\Lambda \Omega$] and [51$\Lambda \Omega$] is the same
in both calculations. The [505 11/2] level is considerably lower in
our case, which is due to differences between the single-particle potentials
in the two models.

\subsection{The microscopic origin of the asymmetry of the fission saddle in rare-earth nuclei}
The case made in Ref.\ \cite{gustafsson71:a} that couplings between levels
of types [41$\Lambda \Omega$] and  [51$\Lambda \Omega$] are the origin
of the fission saddle asymmetry stimulated the discussions,
both those directly  related to fission for example
\cite{mosel71:a,moller74:a} but also
those related to more fundamental properties
\cite{brack05:a,arita16:a,magner16:a}.
However,  discussions in Ref.\ \cite{gustafsson71:a}
of actinide fission were what we sometimes
refer to as postdictions (or postexplanations). But, the paper uses the
insight  gained from the actinides in a less frequent way,
namely to make predictions. In Ref.\ \cite{gustafsson71:a} it is stated
that ``{\it A corresponding situation is expected for nuclei that are one
  shell lighter in neutron number in which case the orbitals}
[$303\,5/2$], [$301\,3/2$], [$301\,1/2$]
{\it should be involved and couple strongly
  \dots to} [$413\,5/2$], [$411\,3/2$] {\it and} [$411\,1/2$]''. And later
in the manuscript 
rare-earth nuclei are mentioned as where these levels may
be suitably positioned to 
lead to asymmetric saddle points. The study also had the insight to
state:
{\it Here the barrier extends to much larger distortions
where the description in terms of solely $\epsilon_2$
and $\epsilon_4$ is far less satisfactory. Then the more
general parameterization suggested by other
groups working in this field along similar lines
is clearly needed [9]}. This reference number points to four different
publications, one of which uses the three-quadratic-surface parameterization
we now use in our investigation here and in many other fission studies.

We have now studied in detail
if the  microscopic mechanism behind the saddle asymmetries
in the sub-Pb region is the simple one predicted
in 1971 \cite{gustafsson71:a}, specifically if the the postulated
involvement of the specific levels mentioned does occur.
We have selected for our
study $^{186}$Pt which is located in the center of the calculated,
(by use of 40-year later, quantitative  models) 
new region of asymmetry \cite{moller15:b}, see the figure from
that paper which we reproduce here as our Fig.\ \ref{asymreg}.
We see that the prediction of Ref.\ \cite{gustafsson71:a} is right
on the mark. For the neutron number $N=108$ of $^{186}_{\phantom{0}78}$Pt
there is a high level density at $\beta_3=0$ but a low density at
$\beta_3=0.08$ which is the calculated value at the asymmetric saddle point,
indicated by a gray bar in the figure. Other observations we can make
is that the saddle point of $^{186}_{\phantom{0}78}$Pt is more elongated
($\beta_2=1.31$) than that of $^{240}_{\phantom{0}94}$Pu ($\beta_2 = 0.85$)
as pointed out in Ref.\ \cite{gustafsson71:a}. The larger asymmetry
($\beta_3= 0.175$) at the saddle point of $^{240}_{\phantom{0}94}$Pu
compared to the smaller ($\beta_3=0.08$) for $^{186}_{\phantom{0}78}$Pt
is also in qualitative
agreement with what can be expected from experiment (about 140/96 for
$^{240}_{\phantom{0}94}$Pu and  100/80 for $^{180}_{\phantom{0}80}$Hg, 
the nearby $^{186}_{\phantom{0}78}$Pt has not been measured yet). More
quantitative methods are used in Refs.\
\cite{moller01:a,moller09:a,moller12:a}. Now we may ask,
since we showed that
the microscopic mechanism between the fission saddle asymmetry
is the same in the actinide region and in the rare-earth region:
is the asymmetry of $^{180}_{\phantom{0}80}$Hg fission ``a new type
of asymmetry''. To try to answer this question we need to discuss
if and how the saddle asymmetry may be related to the final asymmetry
after fragment separation.

\section{The relation between saddle-point asymmetry and fragment asymmetry}
\begin{figure}[t!]
  \includegraphics[width=\linewidth]{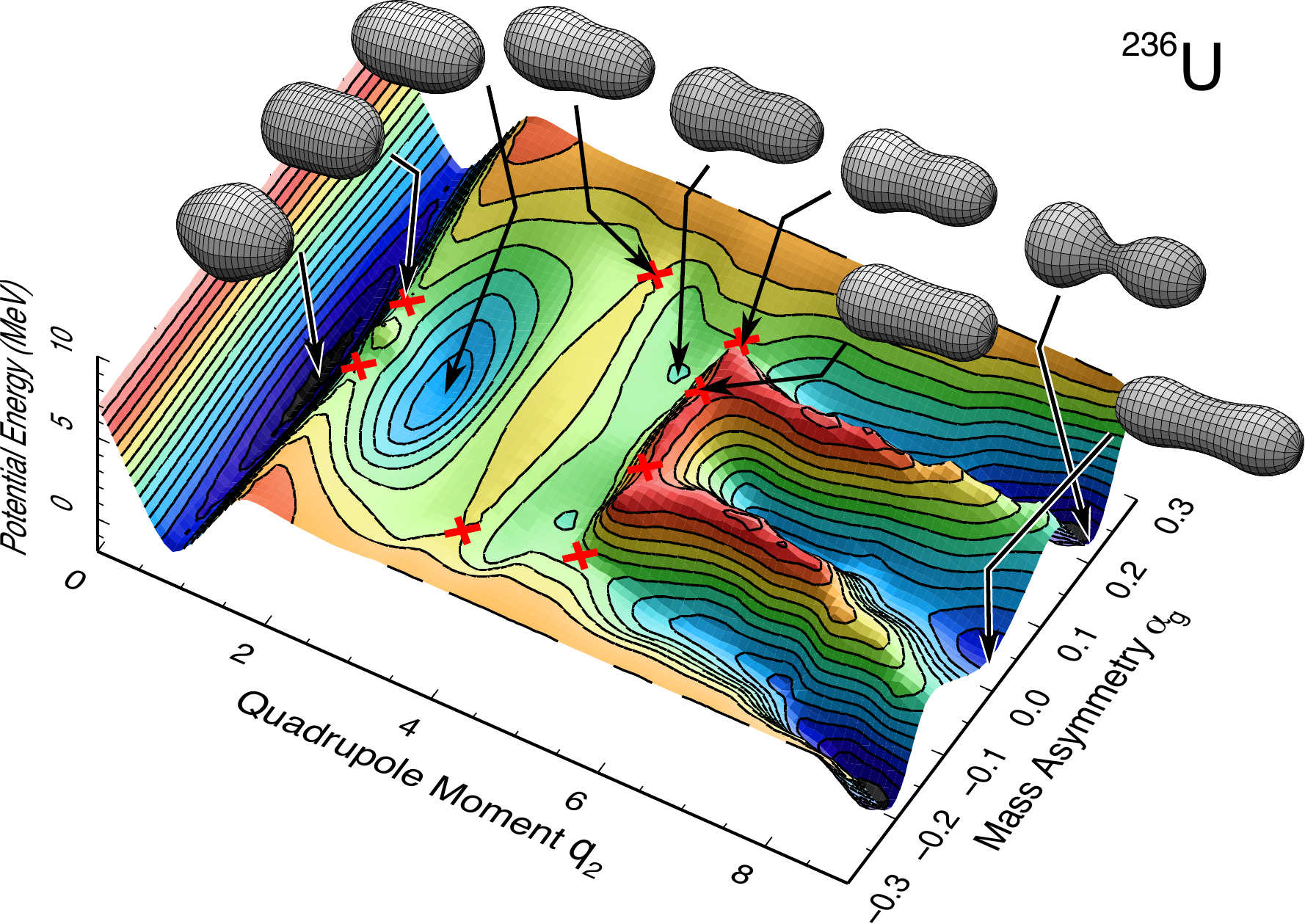}
  \caption{Two-dimensional potential-energy surface
for $^{236}$U which shows some essential features of the full 5D potential-energy
surface. Two crossed (red) lines show the locations of some
saddle points. Note in particular that the valley across the asymmetric
saddle continues to the largest $q_2$ shown, where the nucleus is on
the verge to into two fragments. From \protect\cite{ichikawa12:a}.}
\label{u3d}
\end{figure}

An early calculation of the potential energy versus shape from saddle
to scission for $^{236}$U was the study
in Ref.\ \cite{moller74:a}. In this study the
nuclear shape at the saddle is indeed asymmetric but for for more elongated
shapes symmetry seems reestablished, see Fig. 6 in \cite{moller74:a}.
However, due to computational limitations at
the time, (the figure took 28 hours on a CDC 6600 to calculate) the
figure was based on only 175 different shapes. It was speculated
at the time that this switchback to symmetry after
the fission saddle point might be due to the limited
set of shapes that were (by necessity) considered. About 30 years
later, calculations of  potential-energy surfaces based
on five shape degrees of freedom and up to 5 000 000 different shapes
showed that the saddle asymmetry is indeed preserved (in the form
of a deep valley with constant asymmetry)  in many actinides as they
evolve towards
scission see for example figures
in Refs.\ \cite{moller01:a,moller09:a,ichikawa12:a}. It was found
earlier that for actinides
the final fragment shell structure, such as gaps at
magic numbers, persist to smaller deformations, even to some degree
to elongations corresponding to the second minimum, see Figs.\ 2 and 7
in Ref.\ \cite{moller87:c} and Fig.\ 39 in Ref.\ \cite{moller89:a}.
In Fig.\ \ref{u3d} we show how this level structure affects
a calculated potential-energy surface
by presenting, schematically,
results for $^{236}$U versus elongation $q_2$ and
asymmetry $\alpha_{\rm g}$. The picture is constructed with the aim
to show ``essential features'' of the full 5D potential energy.
Although 2D it is based on our analysis of
the full 5D calculation. We are often asked,
can you not present such surfaces for all your nuclei ``so we can
understand things''. We need to point out again, that one cannot
accurately depict many features of the full 5D space in a 2D
plot like this for many reasons. To ``understand things''
one has to analyze the full 5D space.
One technique often used for reducing the 5D
potential to 2D is to ``minimize'' with respect to the other shape
degrees of freedom. But, if there are several minima  versus
the additional three shape coordinates there is no well-defined way
of doing this minimization. Furthermore the ridge between the asymmetric
valley and the symmetric valley in Fig.\ \ref{u3d} may have
a {\it larger} asymmetry than the bottom of the asymmetric valley
and then it is completely impossible to display that in 2D.
\begin{figure}[t]
  \includegraphics[width=\linewidth]{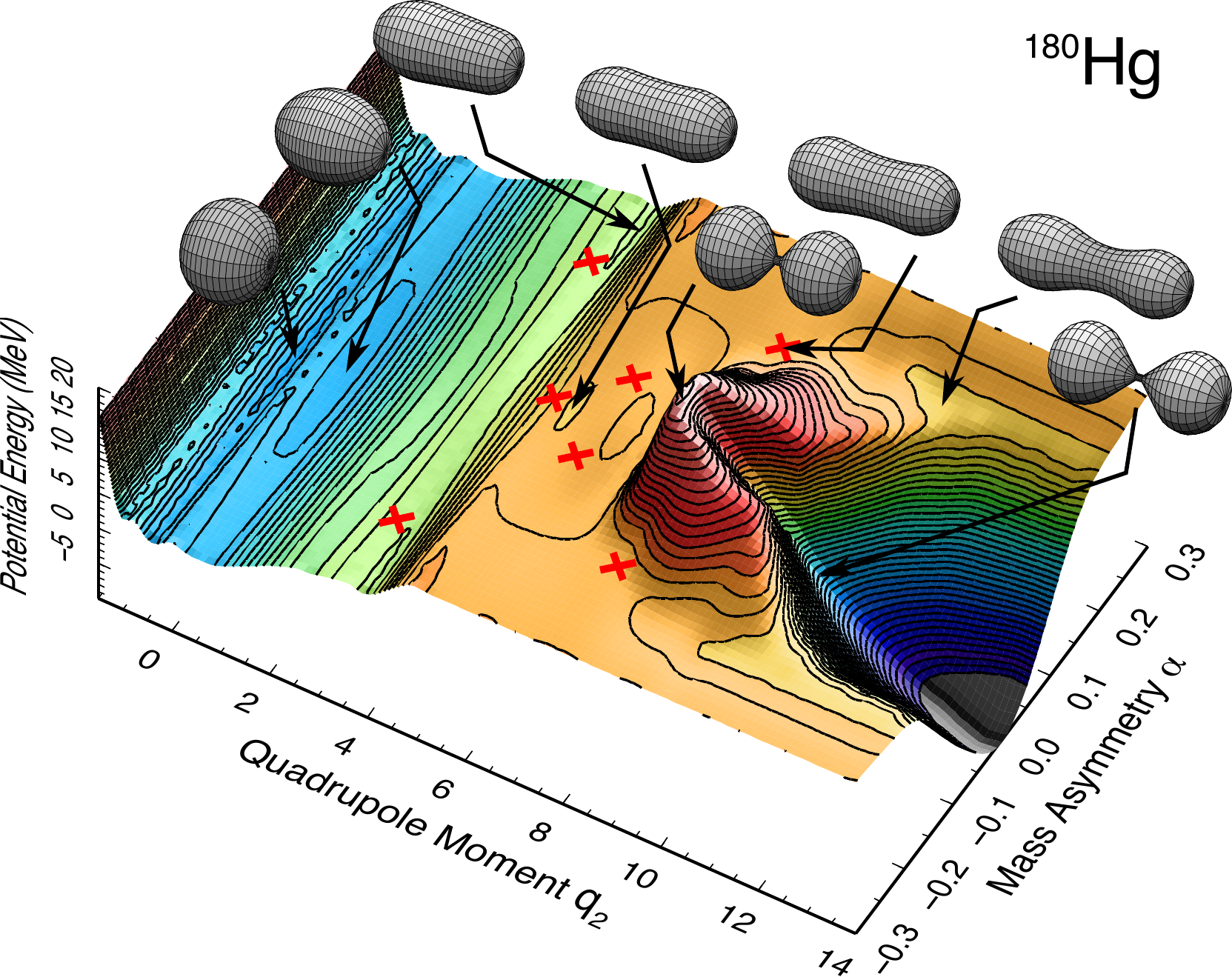}
  \caption{Two-dimensional potential-energy surface
    for $^{180}$Hg which shows some essential features of the full 5D
    potential-energy
surface. Two crossed (red) lines show the locations of some
saddle points. From \protect\cite{ichikawa12:a}.}
\label{hg3d}
\end{figure}
Additional issues are discussed in Ref.\ \cite{moller09:a}.
In our case here the ridge asymmetry is smaller than
that of the asymmetric valley. However, the weakly developed 
saddle from the third minimum into the symmetric valley has a larger
asymmetry than the saddle leading from the third minimum into the asymmetric
valley. Both these saddles are indicated by arrows originating in
the corresponding shapes. Therefore to make it possible to
present 2D figures we improvivsed and just changed the asymmetry of
the saddle leading from the third minimum into the symmetric valley
to a smaller value. This does not change the main features
of the potential energy surface or our conclusions here, namely that
there is a deep, well-developed asymmetric
valley extending continuously from the outer fission saddle to
where the figure ends, corresponding to fragment separation.

In Fig.\ \ref{hg3d} we show schematically in 2D the
essential features of the 5D potential-energy surface of $^{180}$Hg.
Here there is just one symmetric valley at large distortions. The fission
saddle corresponds to asymmetric shapes. Why does this asymmetry
then apparently persist in the final fragment mass split? In Ref.\
\cite{andreyev10:a} it is argued
that already at the elongated saddle shape (much more so than
in the actinide region) a well developed neck is  present. Furthermore
at the very low excitation energy of this experiment,
the mountain and the small ridge  extending
from this mountain to larger elongation  prevents the nucleus from
accessing the symmetric valley until the neck is so small that
the nucleus is separating into fragments before dropping into
the symmetric valley, which is actually a ``fusion'' valley. The differences
between fission and fusion valleys are
discussed in more detail in, for example, Refs.\
\cite{moller76:a,moller87:c,ichikawa05:a,moller09:a}

\section{Summary}
As discussed above many of the theories about the origin of the asymmetry
of fission fragment mass distributions were intuitive,
and not always anchored in a well specified approach
that could routinely be applied to any fissioning system.
The study in Ref.\ \cite{gustafsson71:a} was specific
in several respects: 1) that specific levels were responsible
for the instability of the second saddle to mass-asymmetric
shape deformations, and 2) that a similar situation would
occur in the so-called rare-earth region. It was implicitly assumed
that saddle asymmetry would lead to fragment asymmetry
based on the correlations between calculated saddle properties
and experimental fission yield data known at the time.

Now asymmetry has been observed in the rare-earth regions,
most clearly for $^{180}$Hg. We have shown here that also the mechanism
behind this asymmetry is the one predicted in 1971 \cite{gustafsson71:a},
so the prediction was not just a guess, the physics background
is the one discussed in 1971.

Ideally in comparing ``theory'' to experiment it is desirable
to have a complete model specification, possibly an associated computer
code that does not change in step with new experiments. In
applications to fission fragment yields it should be possible to
routinely apply such a model by simply providing 
the proton and neutron number of the fissioning system and its excitation
energy, run the code and obtain a calculated yield. Until recently no such model
existed. However, soon after the $^{180}$Hg experiment \cite{andreyev10:a}
such a model was developed, namely the Brownian shape-motion model
which implements a random walk on previously calculated 5D potential-energy
surfaces. Details of the model, extensive tests with respect
to experimental data, and sensitivity studies are in Refs.\
\cite{randrup11:a,randrup11:b,randrup13:a}. In this quantitative model
the most likely mass split in low-energy $^{180}$Hg is 104.4/75.6
\cite{moller12:a}; the experimental result is 100/80 \cite{andreyev10:a}.
This model predicts an extended region of fission fragment asymmetry
in the rare-earth region. The 50 year old observation that
interactions between specific single-particle states
would lead to asymmetric fission saddle points and related
asymmetric fragment mass splits hints at an extended, contiguous
region of asymmetry in the rare-earth region, since similar
interactions lead to an extended region of asymmetry
in the actinide region. Therefore,  there are two mutually
supporting results showing an extended
region of asymmetric fission in the rare-earth
region, predictions that can be tested further by experiments.

Our experimental colleagues have repeatedly asked us for
``a simple explanation'' why the $^{180}$Hg fission saddle is asymmetric.
In the words of Ref.\ \cite{gustafsson71:a}
in their conclusions, this has now been accomplished:\\
{\it This analysis based on the \underline{simple}
coupling rules of the asymptotic wave functions
thus appears to give a \underline{simple} understanding of
and strong support to the conclusions reached in
the calculations of ref.\ \cite{moller70:a}} (namely that the
actinide saddle-points are asymmetric; the same argument obviously
carries over to the rare earth region).

This work was carried out under the
auspices of the NNSA of the U.S. Department of Energy at Los Alamos
National Laboratory under Contract No.\ DE-AC52-06NA25396.


\end{document}